\documentclass[manuscript]{aastex}

\shorttitle{What is the redshift of the $\gamma-ray$ BL Lac source S4 0954+65?}
\shortauthors{M. Landoni}

\begin{document}


\title{What is the redshift of the $\gamma-ray$ BL Lac source S4 0954+65?}


\author{M. Landoni}
\affil{INAF - Istituto Nazionale di Astrofisica, Osservatorio Astronomico di Brera. Via E. Bianchi 46 I-23807 Merate (LC) - ITALY }
\email{marco.landoni@brera.inaf.it}
\author{R. Falomo}
\affil{INAF - Istituto Nazionale di Astrofisica, Osservatorio Astronomico di Padova. Vicolo dell'Osservatorio 5 I-35122 Padova (PD)  - ITALY }
\author{A. Treves}
\affil{Universita' degli Studi dell'Insubria, Via Valleggio 11  I-22100  - ITALY }
\author{R. Scarpa}
\affil{Instituto de Astrof'sica de Canarias, C/O Via Lactea, s/n E38205 - La Laguna (Tenerife). Espana}
\author{D. Reverte Pay\'a}
\affil{Instituto de Astrof'sica de Canarias, C/O Via Lactea, s/n E38205 - La Laguna (Tenerife). Espana}

%
%
%
%


\begin{abstract}
High signal-to-noise ratio spectroscopic observations of the BL Lac
object S4 0954+65 at the alleged redshift $z=0.367$ are presented. This source was detected at gamma
frequencies by MAGIC (TeV) and FERMI (GeV) telescopes during a
remarkable outburst that occurred in February 2015, making the determination
of its distance particularly relevant for our understanding of the 
properties of the Extragalactic Background Light.  Contrary to previous reports
on the redshift, we found that the optical
spectrum is featureless at an equivalent width limit of
$\sim 0.1$\AA. A critical analysis of the existing observations indicates that the
redshift is still unknown. Based on the new data we estimate a lower limit to the redshift at 
$z \geq 0.45$.
\end{abstract}


\keywords{S4 0954+65, BL Lacerate objects, redshift, Extragalactic Background Light}



\section{Introduction}
BL Lac objects are extragalactic radio sources hosted in massive elliptical galaxies and are characterised by strong non-thermal synchrotron emission 
dominated by a relativistic jet closely aligned with the line of sight. From the spectroscopic point of view they exhibit a quasi-featureless optical spectrum that makes the detection of
any emission/absorption line extremely difficult. For this reason, the redshift $z$ of many BL Lacs is still
unknown or highly uncertain.  
In particular, the quasi-featureless optical spectrum of BL Lacs is one of the main characteristic for this class of objects. This peculiarity made them rather elusive. The difficulty in detecting features has the direct consequence of not being able to measure their distance. Specifically, for low redshift targets the optical spectra with adequate S/N ratio allow one to detect the absorptions lines from their host galaxies (typically a massive elliptical galaxy) while for higher redshift objects and very luminous (beamed) nuclei the situation is more challenging. In fact, only the availability of very high S/N ratio spectra and relatively high spectral resolution could reveal faint features (see e.g. \cite{land13, land14, sba06}). In the most extreme cases where the nucleus-to-host ratio is severe (see \cite{land14}), only intervening absorption features can be detected in the optical spectra and thus only sound lower limits to the redshift can be assessed.
Nevertheless, the knowledge of the redshift is crucial not only to assess their cosmological role and
evolution, which appears to be controversial due to redshift incompleteness (e.g. \cite{ajello}), but also to properly model their emission mechanism and
energetics (see e.g \cite{fal14} and references therein).  Moreover, BL Lacs (and blazars) are the dominant population of extragalactic sources detected at the highest energies in the $\gamma$-ray sky (e.g
\cite{massaro15} and Fermi LAT Collaboration 3FGL catalog). 
In many cases they are also detected at Very High
Energy (VHE) with Cherenkov telescopes (e.g. MAGIC, Veritas and
HESS, see the Chicago TevCat catalog from \cite{tevcat}\footnote{http://tevcat.uchicago.edu}).  It is consequently of the outmost relevance to know the
distance of these sources to understand how extremely high
energy photons propagate through space and interact with the Extragalactic Background Light \citep{fran08,dom}.

In this paper, we focus on the prototypical BL Lac object S4
0954+65 at the alleged redshift $z = 0.367$ that recently experienced a dramatic flare in the optical and
near-IR band, and was detected for the first time at VHE band by the MAGIC telescope \citep{mir15}.
We present new optical spectroscopy obtained with the 10.4 m Gran Telescopio CANARIAS
(GTC) aiming to firmly determine the redshift of the source.

\section{S4 0954+65}

This source was classified as a BL Lac object by \cite{wal84} and exhibits
all the properties of its class. In fact, in the optical band it is
strongly variable with R apparent magnitudes usually ranging between
15 and 17 (e.g. \cite{rait99}). During a flare which occurred in
2011, it brightened by 0.7 mags in 7 hours \citep{mor14}. Linear polarisation is also
strongly variable and can be as high as 20\%, with large changes of
polarisation angle \citep{mor14}.
The radio morphology is rather complex with jet-like structures
appearing from pc to kpc scales. Superluminal moving components have been reported with
apparent speeds up to $\sim$19$c$  \citep{gab2000, kud10, mor14}). The object was studied also in the X-rays band \citep{perl06,resc09} and it was one
of the first detected extragalactic $\gamma$-ray sources \citep{egret,fermi}.
 
The first attempt to determine the redshift of this source was done by \cite{law86} (L86) who proposed $z = 0.367$. Further optical spectroscopy was then
obtained by \cite{stickel93} (S93) confirming this redshift. However, the two determinations of the redshift are mainly based on different spectral features casting some doubts on the soundness of the proposed redshift. 
Moreover, we note that, in spite of the low z, both ground based imaging (S93) and HST observations \citep{scarpa2000, urry2000} failed to detect the underlying diffuse emission from the host galaxy, suggesting a higher redshift for this source. \\
In February 2015 the target exhibited a dramatic optical activity reaching R $\sim$ 13 (\cite{spi15, bach15} and references therein).  This motivated a TeV observation with the Cherenkov telescope MAGIC, which detected the source with 5$\sigma$ significance \citep{mir15}. A simultaneous high, hard state was reported by FERMI Team \citep{ojha}.

\section{Observations and data reduction}

Medium resolution (R = 1000) optical spectra were gathered at the GRANTECAN 10.4 m
telescope located at the Roque de Los Muchachos observatory, La Palma, Spain. 
Data were obtained on the night of February 28th, 2015. 
The telescope was equipped with the Optical System for Imaging and
low-intermediate-Resolution Integrated Spectroscopy (OSIRIS, \cite{cepa}). We observed the object with two grisms (R1000B and R1000R) in
order to ensure a large spectral coverage (from $\sim$ 0.42 to
1$\mu$m) adopting a slit of 1$^{\prime\prime}$. For each grism, we
secured three individual exposures of 150s each, in order to optimise
cosmic-ray rejection and cosmetic cleaning. The seeing during
the observations was $\sim$ 1$^{\prime\prime}$ and the sky condition was
clear. The accuracy of the wavelength solution (assessed through calibration lamps, sky emission lines and standard star spectra) is $\sim$ 0.1$\textrm{\AA}$.
At the time of observations the R (AB) magnitude of the source
was 15.5 $\pm$ 0.2, as derived from a short exposure acquisition image\footnote{A finding chart with a reasonable field of view is available at \texttt{https://ned.ipac.caltech.edu/dss/HB89\_0954+658.gif} and can be useful for those who want to reproduce the observation}. This magnitude is much fainter ($\sim$ 2.5 mags) than the maximum values detected during the flare episode described above.  Data
reduction was carried out using IRAF\footnote{ IRAF (Image Reduction
  and Analysis Facility) is distributed by the National Optical
  Astronomy Observatories, which are operated by the Association of
  Universities for Research in Astronomy, Inc., under cooperative
  agreement with the National Science Foundation.}  adopting standard
procedures for long slit spectroscopy. The S/N ratio of various bands of the
spectrum ranges between $\sim$50 at the bluest wavelength
(4000-6000 $\textrm{\AA}$), $\sim$120 in the central region (6000-7500 $\textrm{\AA}$) and
$\sim$90 in the reddest range. The optical spectrum is shown in Fig. 1.  

\begin{figure*}
\includegraphics[scale=0.70]{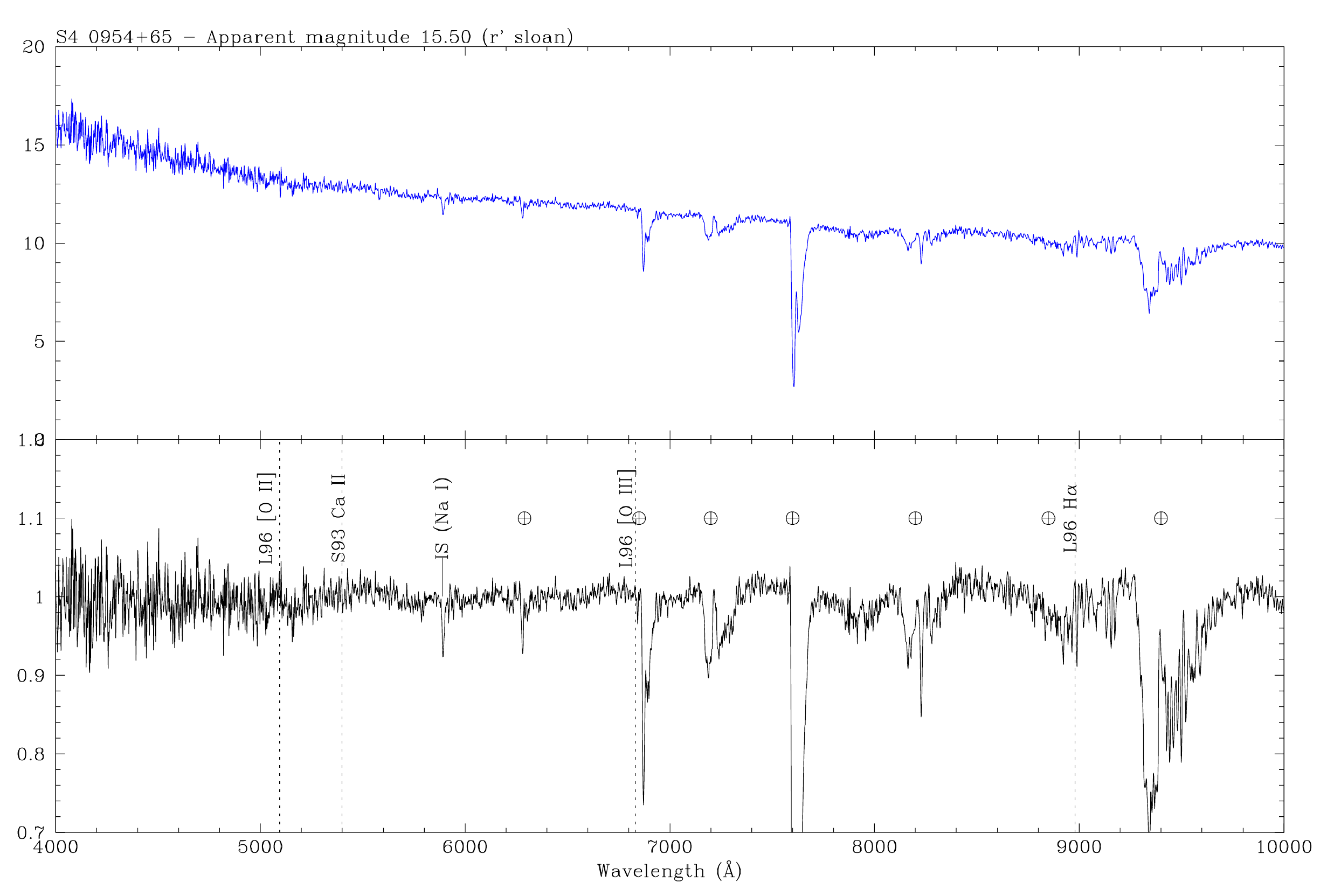}
\caption{S4 0954+65 optical spectrum obtained with OSIRIS at GTC. Upper
  panel: flux calibrated spectrum of the source; Bottom panel:
  normalised spectrum. Interstellar absorption features are marked by ISM while telluric bands are marked by $\oplus$. The position of features claimed by Stickel et a 93 (S93) and Lawrence et al 1996 (L96) are reported as vertical dotted lines (see text).}

\end{figure*}

\section{Results}

The spectrum of S4 0954+65 does not exhibit any intrinsic feature and the continuum is well described by a power law of the form $\lambda^{-\alpha}$ with spectral index $\alpha
\sim$ 0.9. This result is comparabile with that proposed by \cite{law96} and consistent with the mean value of optical-nearIR spectral index for BL Lac objects (see e.g. \cite{pian, land13}). In addition to the prominent telluric absorption bands, we detect the interstellar absorption of ISM Na I with equivalent width (EW) $\sim 0.80 \textrm{\AA}$. We
then calculated the minimum equivalent width $EW_{min}$ detectable on the
spectrum in various bands, adopting the procedure described in
\cite{sba06}. In particular, we evaluated the EW
on bins of the size of the resolution element in various regions of the spectrum excluding the 
telluric structures. We assume as $EW_{min}$ the 2-$\sigma$ deviation from the mean
of the average of the distribution of the EWs obtained in each bin. We found that the EW$_{min}$ is
$\sim$ 0.15$\textrm{\AA}$ in the region 4000-6000$\textrm{\AA}$,  $\sim$ 0.1$\textrm{\AA}$ between
6000-7500$\textrm{\AA}$ and $\sim$0.2$\textrm{\AA}$ in the reddest part.

The result from our spectrum
contrasts with previous claims of detection of intrinsic emission and
absorption features. Specifically, \cite{law96} (L96), on the basis of
spectrum obtained in 1986 (R $\sim$ 400), reported the detection of
three weak narrow emission lines identified as [O II]
($\lambda_{rest}$ 3727, EW$_{rest}$ 0.5 $\textrm{\AA}$) and [O III]
($\lambda_{rest}$ 4959, EW$_{rest}$ 0.3 $\textrm{\AA}$;
$\lambda_{rest}$ 5007, EW$_{rest}$ 0.7 $\textrm{\AA}$) at redshift $z
= 0.3668$. In addition, a broad emission line detected at the edge of
their observed spectrum (EW 2.6 $\textrm{\AA}$) was ascribed to
H$\alpha$ emission at about 8990 $\textrm{\AA}$ in the observer frame. A low resolution optical spectrum was obtained by S93 finding the emission line ascribed to  [O II] $\lambda$3727 (EW 0.3 
$\textrm{\AA}$) but not confirming the [O III] emission line reported by L96. Note that EW of the [O III] line detected by L96 was greater than that of [O II].
On the other hand, S93 found also an
absorption doublet at $\sim$ 5400 $\textrm{\AA}$ ascribed to Ca II
$\lambda\lambda$ 3933-3968 from the host galaxy (not detected by L96 although the
source was at similar flux level). 

In our new spectrum we cannot
confirm any of the previous aforementioned absorption or emission
lines. In particular, the [O II] $\lambda$3727 that would be observed at 5094
$\textrm{\AA}$ at $z = 0.367$ is not detected at the limit of EW$_{min} \geq $ 0.15
$\textrm{\AA}$ (see Figure 1). The [O III] $\lambda\lambda$ 4959,5007 narrow features that would appear at 6804$\textrm{\AA}$, very close to the prominent O$_2$
telluric absorption band, are again not detected at EW$_{min}$ $\geq$ 0.1$\textrm{\AA}$. Moreover, we do not detect neither the
absorption features of Ca II claimed by S93 (down to EW$_{min}$ $\geq$ 0.15
$\textrm{\AA}$), or the broad
emission at $\sim$ 8990 $\textrm{\AA}$ proposed by L96 (heavely contaminated by the strong H$_2$O telluric band). Note also that this
feature at the proposed redshift $z = 0.3668$ corresponds to a rest frame
emission at 6580 (see Table 35 in L96), inconsistent
with H$\alpha$ ($\lambda$ 6563).

It is worth noting that the continuum level during our observation was a
factor of $\sim$ 2-3 higher than that reported in S93 and L96 (see e.g. magnitude reported in Table 1 in S93).  Therefore, if the claimed detected features were unchanged in
intensity, their EWs should be lowered by the same
factor. Considering the dilution due to the enhanced continuum, we
should have detected the [O III] $\lambda$ 5007 at EW $\sim$ 0.3 $\textrm{\AA}$, 3 times higher than our $EW_{min}$. This consideration also applies for H$\alpha$ emission
while it is not possible to give an estimation for the Ca II lines
(S93) since EWs are not reported. \\

Based on our high SNR spectrum, we can estimate a lower limit to the
redshift following the recipe described in \cite{sba06} and \cite{land14}. Briefly, assuming that the BL Lac is hosted by an average
elliptical galaxy with $M_{R} = -22.9$ it is possible to infer a lower limit of the redshift from the $EW_{min}$ and the apparent magnitude of the source (see Equation 1 in \cite{land14}). 
This procedure applied to the spectrum of S4 0954+658 yields $z \geq 0.45$. This lower limit is consistent with the non
detection of the host galaxy from direct imaging (see S93 and \cite{scarpa2000}).

\section{Conclusions}

We presented a high quality optical spectrum of the bright BL Lac object S4 0954+65 that disputes the previous claimed redshift $z=0.367$ adopted in the last $\sim$ 30 years. 
Based on the new spectroscopic observation we estimate a lower limit to the redshift to $z > 0.45$. 
\\
Nowadays, the population of BL Lacs detected at gamma frequencies at GeV and, in particular, at TeV energies is fast growing. Thanks to their emission at very high energy, they can be used as a natural probe for the characterisation of the properties of the Extragalactic Background Light (see e.g. \cite{fran08,dom}). The knowledge of the redshift of the sources (or a lower limit) is therefore mandatory to these studies. In this context, the case of S4 0954+658 is rather prototypical.
On the basis of a number of observable quantities such as the apparent
magnitude, point like images, featureless spectrum and their $\gamma$-ray emission, S4 0954+65 appears similar
to few other bright BL Lacs objects such as PG 1553+113 (R $\sim$ 14) and
H 1722+119 (R $\sim$ 15, \cite{land14}).
The determination of the redshifts for these extreme sources would
require a major improvement of the spectroscopic capabilities as those
expected to be available with the future European Extremely Large Telescope
(see e.g \cite{land13,land14}). Alternatively, using a complementary approach to this kind of sources, one can detect Lyman-$\alpha$ forest absorptions from UV spectra to derive a redshift limit (c.f HST/COS observation by \cite{dan10}). Second, S4 0954+65 has been detected at a threshold energy of hundred of GeVs (see \cite{mir15}) where the opacity due to $\gamma$-$\gamma$ interaction is strongly dependent on the redshift (about a factor of $\sim$2 between 0.35 and 0.45, see \cite{fran08, dom}). In this framework, the newly available estimation of the lower limit of the source will play an important role, especially in the reconstruction of the VHE tail of the emission spectrum of the source, as traditionally assessed for TeV sources absorbed by EBL interaction \citep{prandini}. In this case, the knowledge of the $z$ is decisive to reconstruct the intrinsic, unabsorbed spectrum and then apply the model of emission mechanisms \citep{ghisellini} to recover the physical state of the source (magnetic field, black hole mass, etc.).

\end{document}